\documentclass[acp, manuscript]{copernicus}

\usepackage{booktabs}
\usepackage{varwidth}
\usepackage{ulem,xcolor,amssymb}

\newcommand{\beq}{\begin{equation}}
\newcommand{\eeq}{\end{equation}}

\begin{document}
\nolinenumbers

\title{The equations of motion for moist atmospheric air}

% \Author[affil]{given_name}{surname}

\Author[1,2]{Anastassia M.}{Makarieva}
\Author[1,2]{Victor G.}{Gorshkov}
\Author[1]{Andrei V.}{Nefiodov}
\Author[3]{Douglas}{Sheil}
\Author[4]{Antonio Donato}{Nobre}
\Author[5]{Peter}{Bunyard}
\Author[6]{Paulo}{Nobre}
\Author[2]{Bai-Lian}{Li}

\affil[1]{Theoretical Physics Division, Petersburg Nuclear Physics Institute, Gatchina  188300, St.~Petersburg, Russia}
\affil[2]{USDA-China MOST Joint Research Center for AgroEcology and Sustainability, University of California, Riverside 92521-0124, USA}
\affil[3]{Faculty of Environmental Sciences and Natural Resource Management, Norwegian University of Life Sciences, \AA s, Norway}
\affil[4]{Centro de Ci\^{e}ncia do Sistema Terrestre INPE, S\~{a}o Jos\'{e} dos Campos, S\~{a}o Paulo  12227-010, Brazil}
\affil[5]{Lawellen Farm, Withiel, Bodmin, Cornwall, PL30 5NW, United Kingdom and University Sergio Arboleda, Bogota, Colombia}
\affil[6]{Center for Weather Forecast and Climate Studies INPE, S\~{a}o Jos\'{e} dos Campos, S\~{a}o Paulo 12227-010, Brazil}

%% The [] brackets identify the author with the corresponding affiliation. 1, 2, 3, etc. should be inserted.

\runningtitle{The equations of motion for moist air}

\runningauthor{Makarieva et al.}

\correspondence{Anastassia Makarieva (ammakarieva@gmail.com)}

\received{}
\pubdiscuss{} %% only important for two-stage journals
\revised{}
\accepted{}
\published{}

%% These dates will be inserted by Copernicus Publications during the typesetting process.

\firstpage{1}

\maketitle

\begin{abstract}
How phase transitions affect  the  motion  of  moist atmospheric  air remains controversial. In the early 2000s two distinct  differential equations of motion were proposed.  Besides   their  contrasting formulations  for the acceleration of condensate, the equations   differ  concerning the presence/absence of a term  equal to the rate of phase transitions multiplied by the  difference  in  velocity between condensate and air. This term was interpreted in the literature as the "reactive motion" associated with condensation.  The reasoning behind this "reactive motion" was that when water vapor condenses and droplets begin to fall the remaining gas must move upwards to conserve momentum.  Here we    show that   the two contrasting formulations imply  distinct assumptions about how  gaseous air  and condensate particles interact. We show that these assumptions cannot be simultaneously applicable to condensation and evaporation.  "Reactive motion" leading to an upward acceleration of  air during condensation does not exist. The "reactive motion" term can be justified for evaporation only; it describes  the  downward  acceleration of air.  We emphasize the difference between the equations of motion (i.e., equations constraining velocity) and those constraining momentum  (i.e., equations of motion and continuity combined). We show that, owing to the imprecise nature of the continuity equations, consideration of total momentum can be misleading and that this led to the "reactive motion" controversy.  Finally, we provide a revised and   generally applicable equation for  the motion of moist air.
\end{abstract}

%\Large
\introduction  %% \introduction[modified heading if necessary]

\label{int}

The equation of motion for moist air in the presence of phase changes remains controversial in the meteorological and multiphase flow
literature \citep{young95,drew98,oo01,ba02,brennen05}. \citet{young95} for example reviewed this subject and highlighted a number of inconsistencies among published treatments. In the atmospheric sciences this problem received attention in the works of \citet{oo01} and \citet{ba02}  (hereafter O01 and B02, respectively).  But rather than resolving the issues these authors offered contrasting equations derived
from first principles. \citet{co11} reviewed the fundamental equations of moist atmospheric dynamics
and noted that the correct way to consider phase changes remained poorly understood.

\begin{figure}[th!]
\vspace{-0.5 cm}
\begin{minipage}[p]{\textwidth}
\centering\includegraphics[width=0.8\textwidth,angle=0,clip]{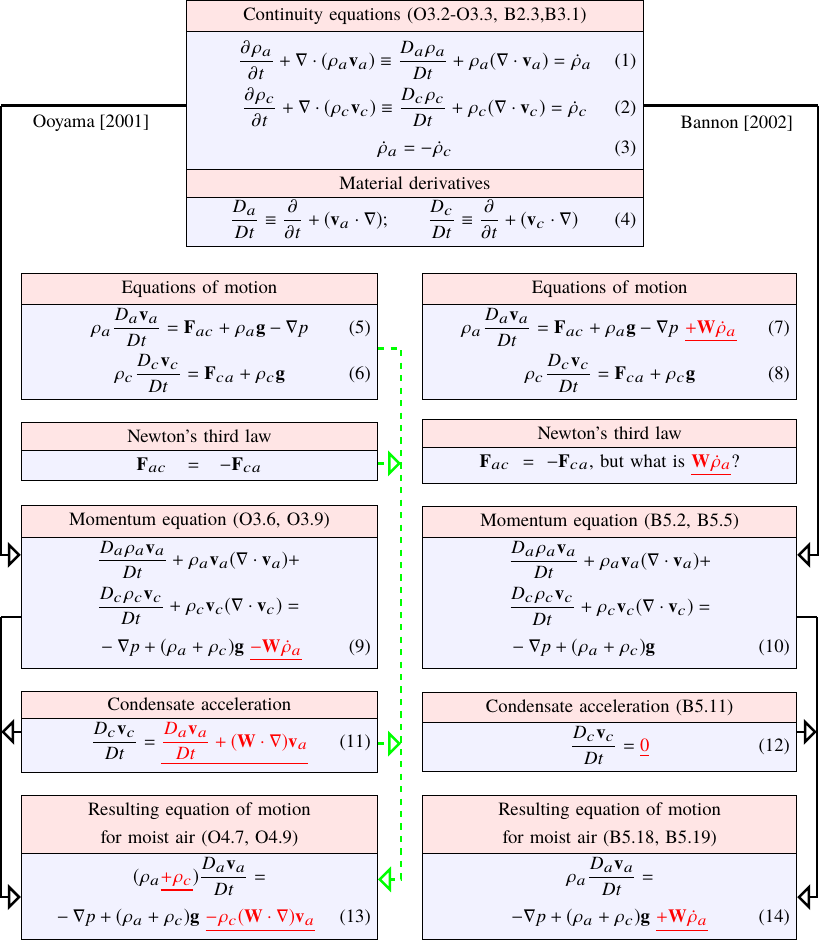}
\end{minipage}
\caption{
Key relationships of O01 (left) and B02 (right)
with differences underlined and highlighted in red.
Subscripts $a$ and $c$ refer to moist air (dry air and water vapor) and condensate, respectively;
$\rho$ (kg~m$^{-3}$) is density; $\dot{\rho}$ (kg~m$^{-3}$~s$^{-1}$) is the rate of phase transitions, $\mathbf{v}$ is velocity; $\mathbf{W} \equiv \mathbf{v}_c - \mathbf{v}_a$;
$p$ is pressure, $\mathbf{g}$ is the acceleration of gravity.
Black solid arrows indicate how the
equation of motion for moist air was derived from the momentum equation:
multiply the continuity equations for air and condensate by, respectively, $\mathbf{v}_a$ and $\mathbf{v}_c$, subtract them from
the momentum equation and use the assumption about condensate acceleration. Green dashed arrows
indicate the derivation from the equations of motion and Newton's third law (not applicable for B02).
Numbers in parentheses marked with "O" and "B" refer to the corresponding equations of O01 and B02.
}
\label{fig1}
\end{figure}

\setcounter{equation}{14}  % reset counter

Here we consider the differences between the formulations of O01 and B02 summarized
in Fig.~\ref{fig1}, Eqs.~(1)-(14). For clarity, and without losing generality, this summary considers
the case when there is only one type of condensate particles (droplets with velocity $\mathbf{v}_c$) and the external forces
acting on moist air are gravity and pressure gradient (Coriolis force and friction are not considered).
The two formulations for moist air, Eqs.~(13) and (14), differ in that B02 includes
the following term
\begin{linenomath*}\begin{equation}\label{Fr}
\mathbf{W} \dot{\rho}_a, \quad   \mathbf{W} \equiv \mathbf{v}_c - \mathbf{v}_a,
\end{equation}\end{linenomath*}
where $\mathbf{W}$ is the difference between condensate velocity $\mathbf{v}_c$ and air velocity $\mathbf{v}_a$,
$\dot{\rho}_a$ is the rate of phase transitions (kg~m$^{-3}$~s$^{-1}$). When condensation occurs, $\dot{\rho}_a < 0$,
see Eq.~(1) in Fig.~\ref{fig1}. (The remaining differences between Eqs.~(13) and (14) will be addressed later.)

\citet[][p. 1972]{ba02} interpreted this term as the "reactive motion" arising during condensation:
as the droplets begin to fall and thus gain a downward velocity, the remaining air gains an upward velocity so that the combined momentum of  air plus droplets is conserved. \citet[][see their Fig. 2.2]{co11} endorsed this interpretation.

However, the "reactive motion" explanation  appears counter to our know\-ledge of  atmospheric processes: indeed, unlike a rocket which accelerates by reactive motion, i.e., by {\it internal} forces  between the rocket and the expelled fuel, droplets upon condensation of water vapor are accelerated downward by a recognised and {\it external} force -- gravity.  (As another example, consider a block of ice melting on a table made of open mesh.  The block does not accelerate upwards as the melt water streams down.)

\citet{ba02} mentioned the disagreement with \citet{oo01} but did not identify either its cause or implications. If, as suggested by \citet{co11}, the formulation of B02 obeys momentum conservation, does the contrasting formulation of O01, used in global atmospheric models \citep{satoh14}, violate it?  While some authors have argued that the "reactive motion" term in Eq.~(14) is usually small  \citep{monteiro07}, \citet{co11} concluded that this term warranted further study.  Irrespective of its magnitude, resolving the discrepancy between the formulations of O01 and B02 is necessary for correct employment of the  fundamental equations of momentum conservation to a moist atmosphere. We undertake such an analysis below considering O01 in Sec.~\ref{O01}, B02 in Sec.~\ref{B02} and summarising the conclusions in Sec.~\ref{conc}.

\section{Derivations of \citet{oo01}}
\label{O01}

\subsection{Equations of motion and momentum equations}

Unlike the meteorological literature where any equation involving air acceleration can apparently be called a "momentum equation", the physics literature is more selective. For example, in the  physical textbook {\it Fluid Mechanics} of   \citet{land} there is no mention of "momentum equations". The equations formulated by Euler  and Navier and Stokes are {\it equations of motion}.  Below the "momentum equations"  denote only those equations that describe  change of  momentum ($\rho_a \mathbf{v}_a$ or $\rho_c \mathbf{v}_c$): for example,  Eqs.~(9) and (10) in Fig.~\ref{fig1}.  Equations that  describe any change of velocity ($\mathbf{v}_a$ or $\mathbf{v}_c$) are referred to as  "equation of motion": for example, Eqs.~(5)-(8)  in Fig.~\ref{fig1}.

Both O01 and B02 begin their derivations from a {\it momentum equation} for the system {\it moist air
plus droplets}, see Eqs.~(9) and (10) in Fig.~\ref{fig1}. Since none of these previous 
authors justify their basic equations, we begin by showing how they could be derived.
The equations of motion for moist air and condensate in their general form can be written as follows:
\begin{linenomath*}\begin{gather}\label{mot}
\rho_a \frac{D_a \mathbf{v}_a}{D t} - \mathbf{F}_a = 0, \\   \label{motc}
\rho_c \frac{D_c \mathbf{v}_c}{D t} - \mathbf{F}_c = 0.
\end{gather}\end{linenomath*}
Here $\mathbf{F}_a$ and $\mathbf{F}_c$ are the volume-specific forces acting on gaseous air and condensate,
respectively, and the material derivatives are defined by Eq.~(4) in Fig.~1.

The use of  Eq.~\eqref{motc}  for condensate particles by both O01 and B02 implies an as\-sump\-tion essential for our
subsequent analysis, namely, that all condensate particles in a local vo\-lu\-me have the same velocity $\mathbf{v}_c$. 
Indeed, applying  Newton's second law to $N$ droplets contained in an  atmospheric  volume $\widetilde{V}_c$ yields 
\begin{linenomath*}\begin{equation}\label{eq17s}
\frac{1}{\widetilde{V}_c} \sum_{i=1}^{N} m_i \frac{d \mathbf{v}_{ci}}{dt} = 
\frac{1}{\widetilde{V}_c} \sum_{i=1}^{N}  \mathbf{f}_{ci}, \quad
\rho_c \equiv \frac{1}{\widetilde{V}_c} \sum_{i=1}^{N} m_i , \quad
\mathbf{F}_{c} \equiv   \frac{1}{\widetilde{V}_c}   \sum_{i=1}^{N}
\mathbf{f}_{ci} , 
\end{equation}\end{linenomath*}
where $m_i$, $\mathbf{v}_{ci}$, and $\mathbf{f}_{ci}$ are the mass, velocity, and force acting on the  $i$-th droplet, respectively. Putting $d\mathbf{v}_{ci}/dt \equiv (\mathbf{v}_{ci} \cdot \nabla) \mathbf{v}_{ci}$   (see Eq.~(4) in Fig.~\ref{fig1}) we find that 
Eq.~\eqref{eq17s} is equivalent to Eq.~\eqref{motc} if and only if $\mathbf{v}_{ci} = \mathbf{v}_{c}$: i.e., all droplets  have the same velocity. If there are  \textit{discrete} types  of condensate particles (ice,  snow, rain) of different size, for each such type a separate equation similar to Eq.~\eqref{motc} is used. As we  will see below, the main theoretical problem is presented by particles that have a \textit{continuous} velocity distribution changing their velocity rapidly in a  local  volume.

Generally, the non-linear equations ~\eqref{mot} and  \eqref{motc} presume the existence of   atmospheric volumes $\widetilde{V}_a$ and $\widetilde{V}_c$  within which the  velocities of, respectively,  air and condensate vary insignificantly  -- such that all the air  within  $\widetilde{V}_a$  and all droplets within $\widetilde{V}_c$ can be assumed  to possess the same velocities  --
$\mathbf{v}_a$  for  air and $\mathbf{v}_c$ for  condensate.

The continuity equations (1)-(3), see Fig.~\ref{fig1}, in O01 and B02 are the same. Let us multiply Eqs.~(1) and (2) by, respectively, $\mathbf{v}_a$ and $\mathbf{v}_c$ and sum them up with the respective equations of motion \eqref{mot} and  \eqref{motc}.
After re-arranging the terms, we obtain the momentum equations for gaseous air and condensate:
\begin{linenomath*}\begin{gather}\label{mom}
\frac{D_a \rho_a \mathbf{v}_a}{Dt} + \rho_a \mathbf{v}_a (\nabla \cdot \mathbf{v}_a) - \mathbf{F}_a - \mathbf{v}_a \dot{\rho}_a
=0,   \\     \label{momc}
\frac{D_c \rho_c \mathbf{v}_c}{Dt} + \rho_c \mathbf{v}_c (\nabla \cdot \mathbf{v}_c) - \mathbf{F}_c - \mathbf{v}_c \dot{\rho}_c
=0 .
\end{gather}\end{linenomath*}
The first two terms in Eqs.~\eqref{mom} and \eqref{momc} represent, respectively, change  of  momentum, per unit volume,  of
a certain amount of air and condensate. Indeed, consider an air parcel with mass $m_a = \rho_a \widetilde{V}_a$ occupying volume $\widetilde{V}_a$.  Its momentum is $m_a \mathbf{v}_a$. Change of momentum, taken per unit air volume, is equal   to
\begin{linenomath*}\begin{equation}\label{dmom}
\frac{1}{\widetilde{V}_a}\frac{d(m_a \mathbf{v}_a)}{dt}  = \frac{d \rho_a \mathbf{v}_a}{dt} + \frac{\rho_a \mathbf{v}_a}{\widetilde{V}_a}   \frac{d \widetilde{V}_a}{dt}.
\end{equation}\end{linenomath*}

Summing Eqs.~(\ref{mom}) and (\ref{momc}) using Eq.~(3) and recalling that $\mathbf{W} \equiv \mathbf{v}_c -
\mathbf{v}_a$ we obtain an equation for the change  of the total momentum, per unit volume, of the system with a constant total mass:
gas occupying volume $\widetilde{V}_a$ and condensate particles occupying volume $\widetilde{V}_c$
that coincide at the considered moment of time, $\widetilde{V}_a = \widetilde{V}_c$:
\begin{linenomath*}\begin{equation}\label{momt}
\frac{D_a \rho_a \mathbf{v}_a}{Dt} + \rho_a \mathbf{v}_a (\nabla \cdot \mathbf{v}_a)
+ \frac{D_c \rho_c \mathbf{v}_c}{Dt} + \rho_c \mathbf{v}_c (\nabla \cdot \mathbf{v}_c) =
\mathbf{F}_a + \mathbf{F}_c - \mathbf{W} \dot{\rho}_a.
\end{equation}\end{linenomath*}
The constancy of mass for this system is dictated by the equality $\dot{\rho}_a(\mathbf{r},t) = -\dot{\rho}_c(\mathbf{r},t)$, see Eq.~(3) in Fig.~\ref{fig1}.  This equation prescribes that gas with velocity $\mathbf{v}_a$ turns into a condensate particle with velocity $\mathbf{v}_c$  locally (i.e., at the same coordinate $\mathbf{r}$) and instantaneously (i.e., at the same time point $t$). Thus, total mass within the considered volumes $\widetilde{V}_a$ and $\widetilde{V}_c$ is conserved.

The difference in formulations of B02 and O01 pertain to their logic in specifying  forces $\mathbf{F}_a$ and $\mathbf{F}_c$. Both authors agree that the external forces to consider are gravity (which acts on both air and condensate) and the macroscopic pressure  gradient (which acts on the air alone but can be neglected for droplets because of their small size).

\citet{oo01} further assumed that whatever forces exist between  air and condensate, they are of equal magnitude but opposite sign by Newton's third law and thus should cancel and vanish  in the sum of $\mathbf{F}_a + \mathbf{F}_c$ in the right-hand part of Eq.~(\ref{momt}), see Eqs.~(5) and (6) in Fig.~\ref{fig1}:
\begin{linenomath*}\begin{equation}\label{Fo}
\mathbf{F}_a = -\nabla p +\rho_a \mathbf{g} + \mathbf{F}_{ac},\quad
\mathbf{F}_c = \rho_c \mathbf{g} + \mathbf{F}_{ca}, \quad   \mathbf{F}_{ac} = -\mathbf{F}_{ca}.
\end{equation}\end{linenomath*}
Here  $\mathbf{F}_{ac}$ and $\mathbf{F}_{ca}$ are the forces exerted on the air by the condensate and on the condensate by the air, respectively; $\mathbf{F}_{ac} = -\mathbf{F}_{ca}$ according to Newton's third law.  Using Eq.~(\ref{Fo}), summing up Eqs.~(\ref{mot})  and (\ref{motc}) (and using the assumption for condensate acceleration (11) to be discussed below) we obtain Ooyama's  Eq.~(13), see green dashed arrows in Fig.~\ref{fig1}. This equation of motion  lacks  a "reactive motion" term.

On the other hand, from   Eq.~(\ref{momt}) we see that even if forces are not specified in the equations of motions \eqref{mot} and \eqref{motc},  the "reactive motion" term  appears in the total momentum equation with the minus sign  (cf. also Eqs.~(7) and (9) in Fig.~\ref{fig1}).  What is the meaning  of this term?

\begin{figure}[t!]
\begin{minipage}[p]{\textwidth}
\centering\includegraphics[width=0.8\textwidth,angle=0,clip]{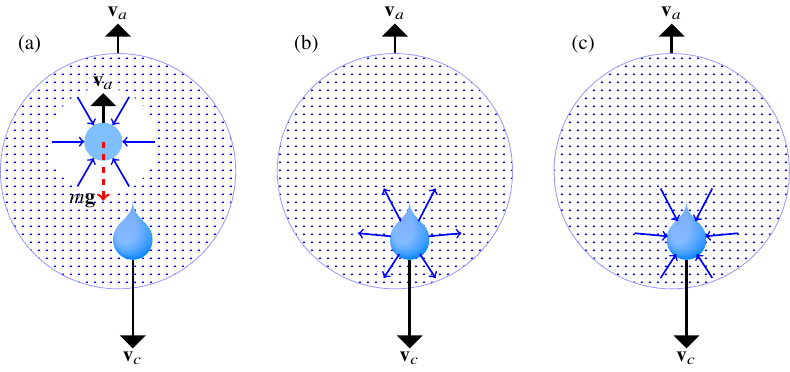}
\end{minipage}
\caption{Different assumptions about interaction between air and condensate particles in the formulations of O01 (a)
and B02 (b), (c). The big circle represents an air volume $\widetilde{V}_a$ moving with velocity $\mathbf{v}_a$
and filled with water vapor (dots). Thin arrows directed to or from the droplets represent condensation (a,c) or evaporation (b), respectively.
In (a), the droplet forms anew from water vapor with velocity $\mathbf{v}_a$ and is accelerated by gravity (red dashed arrow).}
\label{fig2}
\end{figure}

The momentum equation (\ref{momt}), based on  the equations of motion  \eqref{mot} and \eqref{motc},  describes  a system composed of two types of objects: air with velocity $\mathbf{v}_a$ and  droplets  with velocity $\mathbf{v}_c$.  Since the continuity equation demands that $\dot{\rho}_a(\mathbf{r},t) = - \dot{\rho}_c(\mathbf{r},t)$, Eq.~(3) in Fig.~\ref{fig1},  these objects must change their velocity from $\mathbf{v}_a$ to $\mathbf{v}_c$  or vice versa.  However, in reality, such changes cannot be  instantaneous. Accordingly, when phase transitions are occuring in the atmosphere, there  exist  objects with intermediate velocities (between $\mathbf{v}_a$  and  $\mathbf{v}_c$).  As  water vapor condenses into a droplet, the latter has an initital velocity equal to that of local air, Fig.~\ref{fig2}(a).  The droplet then is  accelerated under the action of gravity until  it reaches terminal velocity    $\mathbf{W}$, when gravity is compensated  by the reaction force of the air. For   $W=5$~m~s$^{-1}$ this acceleration takes about  $t_T \sim W/g$  half a second   and ceases at a distance of about  $l_T \sim  W^2/g\sim 3$~m  below the point of condensation.  If $l_T$ is much smaller than the vertical size of the considered atmospheric volume $\widetilde{V}_a$,  most droplets born within this volume will still reside in that volume as they reach terminal velocity.  As noted by \citet{oo01}, as soon as they reach terminal velocity, such droplets are {\it re-classified} as droplets  having velocity $\mathbf{v}_c$, from which point on they obey the equation of motion (\ref{motc}), see also Eq.~(6)  in Fig.~\ref{fig1}.  O01 implicitly assumes that droplets of intermediate  velocity do not interact with either  other droplets  or  air.  Only droplets  with velocity  $\mathbf{v}_c$
exert force $\mathbf{F}_{ac}$ on the air and experience force $\mathbf{F}_{ca}$ from the air.

Thus, in the formulation of \citet{oo01}, $-\mathbf{W} \dot{\rho}_a$ approximates the external force
of gravity that is accelerating the newly formed droplets. Formally, it is as if gas with velocity $\mathbf{v}_a$ disappears from
the system and then instantaneously re-appears as a droplet with velocity $\mathbf{v}_c$ having in the meantime been accelerated by an external force -- gravity. Therefore, the corresponding term $-\mathbf{W} \dot{\rho}_a$ describing this acceleration finds itself in Eq.~(9) among other external forces acting on the system  {\it gas plus droplets with velocity $\mathbf{v}_c$}.  However, since $-\mathbf{W} \dot{\rho}_a$ {\it does not} accelerate either air or  droplets with velocity $\mathbf{v}_c$ (it only accelerates droplets with intermediate velocity),  it is absent from the equations of motion for air and droplets with velocity $\mathbf{v}_c$, i.e., from Eqs.~(5) and (6) in 
Fig.~\ref{fig1}.

\subsection{Acceleration of condensate particles}
\label{acp}

While gravity is responsible for keeping the vertical droplet velocity distinct from the vertical velocity of air, no such forces exist in the horizontal plane. Thus O01 postulated that horizontal velocities $\mathbf{u}_a$ and $\mathbf{u}_c$ of air and droplets coincide,
while vertical velocities $\mathbf{w}_a$ and $\mathbf{w}_c$ differ by $\mathbf{W}$:
\begin{linenomath*}\begin{equation}\label{u}
\mathbf{u}_a = \mathbf{u}_c, \quad   \mathbf{W} \equiv \mathbf{v}_c - \mathbf{v}_a = \mathbf{w}_c - \mathbf{w}_a,
\end{equation}\end{linenomath*}
where $\mathbf{v}_c = \mathbf{u}_c + \mathbf{w}_c$ and $\mathbf{v}_a = \mathbf{u}_a + \mathbf{w}_a$.
Furthermore, \citet[][see  Eq.~(3.10)]{oo01} assumed that $\mathbf{W}$ does not vary along the droplet path:
\begin{linenomath*}\begin{equation}\label{DW}
\frac{D_c \mathbf{W}}{Dt} = 0.
\end{equation}\end{linenomath*}
These two assumptions combined yield Eq.~(11) for droplet acceleration, see Fig.~\ref{fig1}. In the equation of motion~(13) the condensate acceleration plays the role of a drag force imposed by the droplets on the air.

These assumptions are justified when the interaction $\mathbf{F}_{ca}$ between
the air and condensate equalizes velocities more rapidly than they are changed by macroscopic gradients.
Consider the case when the interaction between air and droplets is given by the Stokes force $\mathbf{f}_S$.
The Stokes force is proportional to the velocity difference $\Delta \mathbf{v}$ between air and droplets:
\begin{linenomath*}\begin{equation}\label{fS}
f_S = 3 \pi \rho_a \nu d \Delta v,\quad  a_S \equiv  \frac{f_S}{m},
\quad   \tau_S \equiv  \frac{\Delta  v}{a_S} = \frac{1}{18}\frac{\rho_l}{\rho_a} \frac{d^2}{ \nu} ,
\end{equation}\end{linenomath*}
where $d$ is droplet diameter, $\nu$ is  the kinematic viscosity of air,  $a_S$ is acceleration of a spherical droplet with mass 
$m= \pi \rho_l d^3/6$,  $\rho_l = 10^3$~kg~m$^{-3}$ is the density of liquid water, $\tau_S$ is the time scale at which the Stokes
force  equalizes the velocities  of air and condensate  ($\tau_S$ does not depend on $\mathbf{W}$). 

Let the horizontal air velocity $u_a$ be changed by $\Delta u_a$ over a typical vertical scale  $h \gtrsim 10^2$~m. 
Then condition 
\begin{linenomath*}\begin{equation} \label{eq27}
 \tau_S \ll \frac{h}{W_S}, \quad \tau_S \ll \sqrt{\frac{h}{g}}, \quad W_S \equiv \tau_S g, 
\end{equation}\end{linenomath*}
ensures  that for a droplet falling with terminal velocity $W_S$ we have $|u_c -u_a| \ll | \Delta u_a|$. For small droplets 
with $d < 0.1$~mm obeying the Stokes law  \eqref{fS} we have $\tau_S \sim 0.05$~s assuming $\nu \sim 10^{-5}$~m$^2$~s$^{-1}$
and $\rho_l/\rho_a \sim 10^3$.  The terminal velocity $W$ and the cor\-res\-pon\-ding time scale  $\tau \equiv W/g$  of the largest drops 
with $d \sim 0.6$~cm  are  a factor of  $10^2$  smaller than, respectively, $W_S$  \eqref{eq27} and $\tau_S$  \eqref{fS}  --  due to turbulence  effects \citep{gunkin49}.  Thus  condition \eqref{eq27}  is fulfilled   even for these largest drops when $\tau_S$ is replaced by $\tau \sim 10^{-2} \tau_S$.

These estimates show that the interaction between  air and droplets rapidly  equalizes  their horizontal velocities 
justifying  the assumption  $\mathbf{u}_a = \mathbf{u}_c$  of   O01.  In the vertical plane, due to the presence of gravity,
it aligns droplet velocity with the terminal velocity defined from the balance of the Stokes force  and gravity neglecting the droplet acceleration. This justifies the assumption made by O01 that $\mathbf{W}$ does not change  along the droplet path   and can be parameterized.

\section{Derivations of \citet{ba02}}
\label{B02}

\subsection{Equations of motion and momentum equations}

In the logic of B02,  the terms to appear in the right-hand part of the total momentum equation should only be the external forces (pressure gradient $-\nabla p$ acting on air and gravity $(\rho_a + \rho_c) \mathbf{g}$ acting on air and droplets), see Eq.~(10) in Fig.~\ref{fig1}; they  cannot include anything dependent on the interaction between the droplets and the air since that would be an internal force.

Thus, B02 starts from Eq.~(10), which represents Eq.~(\ref{momt}) without $-\mathbf{W}\dot{\rho}_a$ (Fig.~\ref{fig1}).
Subtracting from Eq.~(10) the continuity equations (1) and (2) multiplied,
respectively, by $\mathbf{v}_a$ and $\mathbf{v}_c$, and using Eq.~(12), which
assumes that droplets do not accelerate, B02 obtains his resulting equation of motion (14) for moist air, see black solid
arrows in Fig.~\ref{fig1}, right column.
This equation contains the "reactive motion" term, which
B02 interpreted as the upward acceleration the air acquires when droplets
with velocity $\mathbf{v}_c$ begin to fall.

However, this derivation appears to conflict with Newton's third law (Fig.~\ref{fig1}).
Indeed, there are only two ways to remove $-\mathbf{W}\dot{\rho}_a$ from the right-hand side
of the total momentum equation: it is to add the same term, but with an opposite sign,
to the formulation of either $\mathbf{F}_a$ or $\mathbf{F}_c$ in the equations of motion (\ref{mot}) or
(\ref{motc}) (i.e., of air or condensate). B02 chooses the former without any stated  justification, cf. Eq.~(\ref{Fo}) of O01:
\begin{linenomath*}\begin{equation}\label{Fb}
\mathbf{F}_a = -\nabla p + \rho_a \mathbf{g} + \mathbf{F}_{ac} + \mathbf{W}\dot{\rho}_a,  \quad
\mathbf{F}_c = \rho_c \mathbf{g} + \mathbf{F}_{ca}.
\end{equation}\end{linenomath*}

Note that if one separately adds $-\mathbf{v}_a \dot{\rho}_a$ to $\mathbf{F}_a$ and
$\mathbf{v}_c \dot{\rho}_a$ to $\mathbf{F}_c$, the resulting equations of motion (\ref{mot}) and (\ref{motc})
will be flawed. Such equations, similar to the equation $dw/dt =-(w/m) dm/dt$ on p.~1972 of B02
aimed to clarify the "reactive motion" during condensation,
violate Galilean invariance by producing different acceleration in different inertial frames of reference.
For a discussion of related errors in the literature see, e.g., \citet{pla92} and \citet{irschik04}.

The question is, if $\mathbf{W} \dot{\rho}_a$ is an internal  force acting between the air and condensate, which is the reason it did not show up in the right-hand part of  the total momentum equation (10), then why doesn't this force obey Newton's third law?
In other words, why isn't this force present with opposite signs in the equations of motion for both condensate and droplets?

This apparent contradiction can be resolved and the true physical meaning of the "reactive motion" term clarified  if we once again take into consideration the existence of objects with intermediate velocity.  In the case of evaporation, Fig.~\ref{fig2}(b), such an object is the water vapor that has just evaporated from a droplet with velocity $\mathbf{v}_c$.  This water vapor initially has velocity $\mathbf{v}_c$ equal to that of the droplet it evaporated from. It then interacts with local air in the volume $\widetilde{V}_a$.  The result of
this interaction is that their velocities equalize. This process is equivalent to an inelastic collision between an  amount of air (water vapor) with velocity $\mathbf{v}_c$  and another amount of air with velocity $\mathbf{v}_a$,
which subsequently  move with  the same velocity.  Since during evaporation $\dot{\rho}_a > 0$, and the droplets have a downward velocity relative to the air,  the "reactive motion" term in Eqs.~(7) and (14) describes the downward acceleration of air as it mixes with water vapor.

For acceleration $\mathbf{a}_v$ of just evaporated  water vapor of mass $m_v$ we can write
\begin{linenomath*}\begin{equation}\label{av}
\frac{m_v \mathbf{a}_v}{\widetilde{V}_a} = \mathbf{F}_{va},\quad   \mathbf{F}_{va}= -\mathbf{F}_{av},\quad   \mathbf{F}_{av} = \mathbf{W} \dot{\rho}_a.
\end{equation}\end{linenomath*}
Here $\mathbf{F}_{va}$ is the force, per unit volume, exerted by the air on the evaporating water vapor;
$\mathbf{F}_{av}$ is, by Newton's third law, the opposite force exerted by the water vapor on the local air.
Since acceleration of this water vapor occurs instantaneously, as dictated by the continunity equation (3),
the steady-state mass $m_v$ of such water vapor with intermediate velocity approaches zero;
their product is finite and equals the "reactive motion" term, Eq.~(\ref{av}). Thus, $\mathbf{W} \dot{\rho}_a$ in Eqs.~(7) and (14) does not violate Newton's third law as it describes interaction of local air not with
the droplets of velocity $\mathbf{v}_c$ but with the  "just evaporated"  water vapor of intermediate velocity, zero mass
and infinite acceleration.

Notably, evaporation does not affect droplet motion, since evaporating water vapor has the same velocity as the droplet from which it evaporates. Thus, during evaporation, the equation of motion for droplets does not include $\mathbf{W} \dot{\rho}_a$.

One could formally consider condensation to occur not by the birth of new drop\-lets, but by the same "inelastic collision"
of water vapor molecules with pre-existing droplets falling at velocity $\mathbf{v}_c$, Fig.~\ref{fig2}(c). In this case  it is the droplet velocity that is affected by interaction with the vapor of different velocity: the droplets  will acquire an upward ac\-ce\-le\-ra\-tion.  The "reactive motion"  term will then be present in the equation of motion for the droplets (8),  but absent from the equation of motion for air (7). The resulting equation of motion for moist air which contains  their sum would be the same. However, such a representation rules out the birth of new droplets anywhere except the only initial level in the atmosphere from which they are falling. In addition, rate of  condensation can be proportional  to the vertical  velocity of the local air, while the number of droplets depends on their size. Therefore, requiring condensation to occur on pre-existing droplets is equivalent to postulating a link between two unrelated parameters. This means that the "reactive motion" term  cannot be justified for condensation.

\subsection{Acceleration of condensate}
\label{acp2}

In the presence of  condensation by "inelastic collision", Fig.~\ref{fig2}(c),   the "reactive motion" term would reside in the equation of motion for drop\-lets. Without taking into account other forces, the reactive force would accelerate  droplets  upwards. Thus, neglecting droplet acceleration, see Eq.~(12), while keeping a "reactive motion" term,  appears contradictory and  unjustified.  With a typical rainfall rate  $P \sim 10^{-3}$~kg m$^{-2}$s$^{-1}$  ($\sim 4$~mm h$^{-1}$) the mean condensation rate in an atmospheric column of height $h \sim 10^3$~m is  $\dot{\rho}_a \sim   - P/h  \sim - 10^{-6}$~kg m$^{-3}$s$^{-1}$.  Meanwhile with $\rho_c \sim 10^{- 4}$~kg m$^{-3}$ and $|\partial v_a/\partial z | \sim  \Delta v_a/h$ with $\Delta v_a \sim 10$~m s$^{-1}$ we have $\rho_c |\partial v_a/\partial z |  \sim  |\dot{\rho}_a|$. This implies that the  "reactive motion" term $\mathbf{W}\dot{\rho}_a$ and condensate acceleration $\rho_c (\mathbf{W} \cdot \nabla) \mathbf{v}_a$,  Eq.~(11),  can be of similar magnitude.

Finally, if all condensation occurred only on pre-existing droplets, the droplets would  grow in size as they fell. Since terminal velocity depends on droplet size,  the assumption of  zero acceleration of droplets adopted in B02 would be invalid.

\section{Conclusions}
\label{conc}

While a valid equation of motion must conform to Galilean invariance and  to fundamental conservation laws, it cannot be derived from those constraints alone.  We have seen that the equation of total momentum  \eqref{momt}  represents a sum of  two equations of motions and two continuity equations each multiplied by the  respective velocity of air or condensate.  Accordingly, the momentum equation  carries no additional information about  system dynamics that is already contained in the  equations of motion and continuity (the latter only 
providing an approximation  since they assume instantaneous velocity changes between air and droplets at phase transitions).  Employing the  total momentum equation without explicit consideration of the equations of motion (\ref{mot}) and (\ref{motc})  resulted in confusion concerning the "reactive motion" term of B02.

For droplets born during condensation the "reactive motion"  term describes their downward acceleration,  Fig.~\ref{fig2}(a).  For evaporation from droplets the term describes the downward acceleration of air,  Fig.~\ref{fig2}(b). For condensation on  pre-existing droplets the term  describes the upward acceleration of droplets, Fig.~\ref{fig2}(c).  In  none of these cases, contrary to previous suggestions, does the "reactive motion"  term describe  the upward acceleration of air. Such a process does not exist.

A generally applicable equation taking these processes into account is (cf. Eqs.~(13) and (14))
\begin{linenomath*}\begin{equation}\label{eq30}
(\rho_a + \rho_c) \frac{D_a \mathbf{v}_a}{D t} = - \nabla p + (\rho_a + \rho_c) \mathbf{g} - 
\rho_c (\mathbf{W} \cdot \nabla) \mathbf{v}_a + \mathbf{W} \dot{\rho}^+_a .
\end{equation}\end{linenomath*}
Here $\dot{\rho}^+_a  > 0$ is the rate of evaporation (e.g., from big drops under the cloud base).  This term will be absent where  condensation occurs ($\dot{\rho}_a <0$). Equation  \eqref{eq30} accounts for the condensate drag neglected  by B02.  As we estimated in Sec.~\ref{acp},  this drag formulated by O01 correctly represents the interaction  between  air and condensate  under most atmospheric conditions.

Phase transitions incur abrupt changes of fluid properties that occur near instantaneously compared to the typical time
scales of air motion.  Here we have considered how the instantaneous formation of condensate  affects the motion  of  moist air.  However, the dynamic effect of phase transitions is not confined to the interaction between air and condensate.  An additional  effect  is the local pressure  perturbation that arises  during  condensation, see Fig.~\ref{fig2}(a). In a hydrostatic atmosphere such perturbations  can cause larger scale  pressure adjustments. These processes can transform potential energy contained in local condensation-induced pressure perturbations into  the   potential energy of   pressure gradients able to drive macroscopic air motions. Constraining this  effect  of condensation  is, in our view, a promising path for future research.

\begin{acknowledgements}
We thank three anonymous referees for  their  constructive comments.  This work is par\-ti\-al\-ly supported by  the University of California Agricultural Experiment Station,  Australian Research Council project DP160102107 and the CNPq/CT-Hidro - GeoClima project Grant~404158/2013-7. 
\end{acknowledgements}

\bibliographystyle{copernicus}
%\bibliography{met-refs}

\begin{thebibliography}{12}
\providecommand{\natexlab}[1]{#1}
\providecommand{\url}[1]{{\tt #1}}
\providecommand{\urlprefix}{URL }
\expandafter\ifx\csname urlstyle\endcsname\relax
  \providecommand{\doi}[1]{doi:\discretionary{}{}{}#1}\else
  \providecommand{\doi}{doi:\discretionary{}{}{}\begingroup
  \urlstyle{rm}\Url}\fi

\bibitem[{Bannon(2002)}]{ba02}
Bannon, P.~R.: Theoretical foundations for models of moist convection, J.
  Atmos. Sci., 59, 1967--1982,
  \doi{10.1175/1520-0469(2002)059<1967:TFFMOM>2.0.CO;2}, 2002.


\bibitem[{Brennen(2005)}]{brennen05}
Brennen, C.~E.: Fundamentals of multiphase flows, Cambridge University Press,
  2005.

\bibitem[{Cotton et~al.(2011)Cotton, Bryan, and van~den Heever}]{co11}
Cotton, W.~R., Bryan, G.~H., and van~den Heever, S.~C.: Fundamental equations
  governing cloud processes, in: Storm and Cloud Dynamics, vol.~99 of {\em
  International Geophysics Series\/}, pp. 15--52, Academic Press, 2 edn., 2011.

\bibitem[{Drew and Passman(1998)}]{drew98}
Drew, D.~A. and Passman, S.~L.: Theory of multicomponent fluids, vol. 135 of
  {\em Applied Mathematical Sciences\/}, Springer, 1998.

\bibitem[{Irschik and Holl(2004)}]{irschik04}
Irschik, H. and Holl, H.~J.: Mechanics of variable-mass systems--{Part 1}:
  {Balance} of mass and linear momentum, Appl. Mech. Rev., 57, 145--160, 2004.

\bibitem[{Gunn and Kinzer(1949)}]{gunkin49}
Gunn, R. and Kinzer, G.~D.: The terminal velocity of  fall for water droplets in stagnant air,
 J. Meteor., 6, 243--248,   \doi{10.1175/1520-0469(1949)006<0243:TTVOFF>2.0.CO;2},  1949.

\bibitem[{Landau and Lifshitz(1987)}]{land}
Landau, L.~D. and Lifshitz, E.~M.: Course of Theoretical Physics. Fluid
  Mechanics, vol.~6, Butterworth-Heinemann, 2 edn., 1987.

\bibitem[{Monteiro and Torlaschi(2007)}]{monteiro07}
Monteiro, E. and Torlaschi, E.: On the dynamics interpretation of the virtual
  temperature, J. Atmos. Sci., 64, 2975--2979, 2007.

\bibitem[{Ooyama(2001)}]{oo01}
Ooyama, K.~V.: A dynamic and thermodynamic foundation for modeling the moist
  atmosphere with parameterized microphysics, J. Atmos. Sci., 58, 2073--2102,
  \doi{10.1175/1520-0469(2001)058<2073:ADATFF>2.0.CO;2}, 2001.

\bibitem[{Plastino and Muzzio(1992)}]{pla92}
Plastino, A.~R. and Muzzio, J.~C.: On the use and abuse of {Newton's} second
  law for variable mass problems, Celestial Mechanics and Dynamical Astronomy,
  53, 227--232, 1992.

\bibitem[{Satoh(2014)}]{satoh14}
Satoh, M.: Atmospheric Circulation Dynamics and General Circulation Models,
  Springer, Heidelberg, 2014.

\bibitem[{Young(1995)}]{young95}
Young, J.~B.: The fundamental equations of gas-droplet multiphase flow, Int. J.
  Multiphase Flow, 21, 175--191, 1995.
  
\end{thebibliography}

\end{document}